\newcommand{\be}{\begin{equation}} \newcommand{\ee}{\end{equation}}
\newcommand{\bea}{\begin{eqnarray}}\newcommand{\eea}{\end{eqnarray}}
\newcommand{\nn}{\nonumber}

\newcommand\Tq{\mbox{Tr}_q}

\documentstyle[12pt]{article}
\begin{document}
\renewcommand{\thefootnote}{\fnsymbol{footnote}}
\begin{titlepage}
{\hfill JINR-E2-94-449 } \\

{\hfill hep-th/9411186}\\

{ November 1994}\vspace{2.5cm}\\

\vspace*{3cm}
\begin{center}

B.M.Zupnik${}\;$\footnote{E-mail: zupnik@thsun1.jinr.dubna.su}\\
${}$Bogoliubov   Laboratory of Theoretical Physics , JINR,
 Head Post Office,\\
P.O.Box 79, 101 000 Moscow, Russia\\

\vspace{2cm}
SOLUTION OF SELF-DUALITY EQUATION IN QUANTUM-GROUP \\
GAUGE   THEORY AND QUANTUM HARMONICS \\

\vspace{2cm}
Submitted to "Physics Letters B"
\vspace{1cm}\\
{\bf Abstract}
\end{center}

We discuss the gauge theory for quantum group $SU_q (2)\times U(1) $ on
 the  quantum Euclidean space. This theory contains three physical gauge
fields and one$\; U(1)-$gauge field with a zero field strength. We
 construct the quantum-group self-duality equation (QGSDE) in terms of
differential forms and  with the help of the field-strength
decomposition. A deformed analog of the BPST-instanton solution is
obtained. We consider a harmonic (twistor) interpretation of QGSDE in
terms of $SU_q (2)/U(1) $ quantum harmonics. The quantum harmonic gauge
 equations are formulated in the framework of a left-covariant 3D
 differential calculus on the quantum group $SU_q (2)$.

\end{titlepage}

\renewcommand{\thefootnote}{\arabic{footnote}}
\setcounter{footnote}0

  An attractive idea of quantum deformations for the gauge theories has
 been considered in the framework of different approaches [1 - 6].
Formally one can discuss independent deformations of basic spaces and
gauge groups and possible correlations between these deformations. We
shall here  consider a gauge theory with identical one-parameter
 deformations of the 4-dimensional Euclidean space and the gauge group
 $SU(2)$. A consistent  formulation of the  gauge theory for the
semisimple quantum group $SU_q (N)$ is unknown to us, so we shall deal
with the  quantum group $U_q (2)=SU_q (2)\times U(1) $. It will be shown
that the $ U(1)$-gauge field can be treated as a field with a zero field
 strength.

   Consider the standard relations between  elements $T_k^i $ of the
quantum $U_q (2)$-matrix \cite{a8}
\be
R T_1 T_2 = T_1 T_2 R\;,\;\;\;\;\;R^2=I+(q-q^{-1})R        \label{A1}
\ee
where $I$ is a unity matrix , $\;R$  is a constant symmetric matrix
 with components $R^{ik}_{lm}(q)\;\;\\ (i , k , l , m =1,\;2)$ and $q $
 is a real  deformation parameter.

It is convenient to use the following covariant representation for a
deformed antisymmetric symbol
\bea
&\varepsilon_{ik}(q)=\sqrt{\;q(ik)}\;\varepsilon_{ik}=-q(ik)
\varepsilon_{ki}(q) & \\ \label{A2}
 q(12)=[q(21)]^{-1}=q,&\;\;\;q(11)=q(22)=1 & \\ \nn
&\varepsilon_{ik}(q)\varepsilon^{kl}(q)=\delta_i^l\;,\;
& \label{A4}
\eea
where   $\varepsilon_{ik}$ is an ordinary
antisymmetric symbol $(\varepsilon_{ik}=\varepsilon^{ki}) $ .

The $R$-matrix  can be written in terms
of projection operators $P^{(+)}$ and $P^{(-)}$\cite{a8}

\bea
& R=q P^{(+)}-q^{-1}P^{(-)}=qI -(q+q^{-1})P^{(-)} & \\ \label{A5}
& P^{(+)}\;+\;P^{(-)}=I\;,&\\ \nn
&(P^{(\pm)})^2 =P^{(\pm)},\;\;\;\; P^{(+)}P^{(-)}=0 &
\label{A6}
\eea
where matrix $P^{(-)}$ has the following components
\be
[P^{(-)}]^{ik}_{lm}=-\frac{q}{1+q^2}
\varepsilon^{ki}(q)\varepsilon_{ml}(q)  \label{A7}
\ee

We shall use also covariant representation for the $\;SU_q (2)$-metric
\be
{\cal D}^i_k (q)=-\varepsilon_{mk}(q)\varepsilon^{mi}(q) \label{A8}
\ee

The basic RTT-relations imply the simple equation
\be
\varepsilon_{ml}(q)T^l_i\;T^m_k = \varepsilon_{ki}(q)\; D(T) \label{A9}
\ee
where $D(T)$ is the quantum determinant \cite{a8}.

A covariant expression for the inverse quantum matrix $ S(T)=T^{-1}$
contains inverse determinant
\be
 S^i_k =\varepsilon_{kl}(q)\;T_j^l\;\varepsilon^{ji}(q)\; D^{-1}(T)
\label{A10}
\ee

We shall use the well-known equations for multiplication of the
transposed matrices
\be
 T^l_i {\cal D}^m_l (q) S^k_m = {\cal D}^k_i (q)  \label{A11}
\ee

The unitarity condition for the matrix $T$ can be formulated with the
help of involution \cite{a8}
\be
T^i_k \rightarrow \overline{T^i_k}= S^k_i \label{A12}
\ee

Let us consider the bicovariant differential calculus on the $U_q(2) $
 group [9 - 12]

\bea
& T_1 dT_2 = R dT_1 T_2 R  & \\ \nn
& D(T) dT = q^2 dT D(T) & \\ \label{A13}
 \left[d,T\right]=dT ,&\;\;\;\;\; \left\{d,dT\right\}=0 & \nn
\eea

Note that the condition $D(T)=1$ is inconsistent in the framework of
this calculus. Consider the relations for the right-invariant
differential forms $ \omega =dT S $\cite{a12}
\bea
& \omega R \omega + R\omega R \omega R =0 & \\ \label{A14}
& T_1 \omega_2 = R \omega_1 R T_1  & \label{A15}
\eea

The quantum trace $\xi$ of the form $\omega$ plays an important role in
this calculus
\bea
 \xi (T)= {\cal D}^k_i (q) \omega^i_k (T) \neq 0,&\;\;\xi^2=0,\;\;\;&
d\xi=0  \\ \nn
 dT=\omega T=(q^2 \lambda)^{-1} [T,\xi] ,&\;\;\;qdD(T)=\xi D(T) &\\
\label{A16}
& d\omega=\omega^2=-(q^2 \lambda)^{-1} \left\{\xi,\omega\right\} & \nn
\eea

The bicovariant calculus makes the basis for consistent formulation of
quantum-group gauge theory in the framework of noncommutative algebra of
differential complexes [5-7]. Consider formally
the quantum  group gauge  matrix     $T^a_b(x)$ defined on some basic
space. Suppose that Eqs(12-15) satisfy locally for each "point" $x$.
Then one can try to construct the $U_q (2)$-connection 1-form
$A^a_b (x)$ which obeys the simplest commutation relation
\be
 A \;R \;A + R \;A \;R \;A \;R\; =\;0  \label{A17}
\ee
Note  that the general relation for $A$ contains a nontrivial right-hand
 side \cite{a7}.

Coaction of the gauge quantum group $U_q (2)$ has the following standard
form:
\bea
& A \rightarrow T(x)\;A\;S(T)\;+\;dT(x)\;S(T) = T\;A\;S\;+\;\omega(T)&\\
\label{A18}
& \alpha = \Tq A \rightarrow \alpha + \xi(T) &\nn
\eea

The restriction $\alpha=0$ is inconsistent with (\ref{A17}),
 but we can use the gauge-covariant relations
\be
\alpha^2=0\;,\;\;\;\;\; \Tq A^2=0 \label{A19}
\ee

It should be stressed that we can choose the zero field-strength
 condition $d\alpha=0$ for the $U(1)$-gauge field \footnote{This
 condition is consistent also for the case of $GL_q(N)$ group}.
 This constraint is gauge invariant and consistent with (\ref{A17}).
 The deformed pure gauge field $\alpha$ can be
decoupled from the set of physical fields in the limit $q=1$. We shall
 further consider the $U_q (2)$-gauge theory with three "physical" gauge
fields and one "zero-mode" $U (1)$ field.

The curvature 2-form is $q$-traceless for this model
\be
F = dA - A^2 ,\;\;\;\;\; \Tq F = 0  \label{A20}
\ee

Quantum deformations of Minkowski and Euclidean 4-dimensional spa\-ces
have been considered in Refs [13-15]. We shall treat the coordinates
$x^i_\alpha $ of $q$-deformed Euclidean space $E_q(4)$ as generators of
 a  noncommutative algebra covariant under the coaction of the quantum
 group $G_q(4)=SU_q^L (2)\times SU_q^R (2)$
\be
 R^{ik}_{lm}x^l_\alpha x^m_\beta = x^i_\gamma x^k_\rho R^{\gamma \rho}_
{\alpha \beta} \label{A21}
\ee
where we use two identical copies of R-matrices (4) for left and
 right $SU_q (2)$-indices.

Coactions of the commuting left and right $SU_q (2)$ groups conserve
(\ref{A21})
\be
x^i_\alpha \rightarrow l^i_k x^k_\beta r^\beta_\alpha  \label{A22}
\ee

The $q$-deformed central Euclidean interval $\tau$ can be constructed by
analogy with the quantum determinant
\be
\tau=\mid x\mid^2 =-\frac{q}{1+q^2}
\varepsilon^{\beta\alpha}(q)\varepsilon_{ki}(q)x^i_\alpha x^k_\beta
\label{A23}
\ee

We do not consider the quantum group structure on $E_q (4)$. It is
convenient to use the following $E_q (4)$ involution
\bea
& \overline{x^i_\alpha}=\varepsilon_{ik}(q) x^k_\beta
\varepsilon^{\beta\alpha}(q)
 =\tau S^\alpha_i (x) & \\ \label{A24}
& \overline{\tau}=\tau\;,\;\;\;\overline{\overline{x^i_\alpha}}=x^i_
\alpha & \nn
\eea

We shall use an analog of the bicovariant $U_q (2)$-calculus (12-15)
for  studying  differential complexes on $E_q (4)$. Consider the
 right-invariant \\ 1-forms
\be
 \omega(x)^i_k = dx^i_\alpha S^\alpha_k (x)  \label{A25}
\ee

Basic 2-forms on $E_q (4)$ can be decomposed with the help of $P^(\pm)$
\\
operators (\ref{A6})
\be
dx^i_\alpha dx^k_\beta= \frac{q}{1+q^2}[\varepsilon_{\beta\alpha}(q)
d^2 x^{ik}\;+\;\varepsilon^{ki}(q)d^2 x_{\alpha \beta}] \label{A26}
\ee

By analogy with the classical case we can treat these two parts as
 self-dual and anti-self-dual 2-forms under the action of a duality
 operator $\ast$ .
It is convenient to rewrite this decomposition in terms of the
right-invariant self-dual and anti-self-dual forms
\bea
& P^{(-)}dx_1 dx_2 P^{(+)}=q P^{(-)}(q^3\omega^2 + \omega\xi)P^{(+)}
x_1 x_2 P^{(+)} & \\ \label{A27}
& P^{(+)}dx_1 dx_2 P^{(-)}=-(1/q)P^{(+)}(\omega R\omega )P^{(-)}\tau=
 (q^{-1} \omega\xi - \omega^2)P^{(-)}\tau &
 \label{A28}
\eea

Let us introduce the simple ansatz for quantum $U_q (2)$ anti-self-dual
gauge fields
\bea
& A^a_b = dx^i_\alpha A^{\alpha a}_{ib}(x) = \omega^a_b (x)f(\tau)\;,
 &\\ \label{A29}
& A^{\alpha a}_{ib}(x)=\delta^a_i S^\alpha_{b}(x)f(\tau)\;, & \nn
\eea
where $f(\tau) $ is a function of the $ q$-interval (\ref{A23}). Note
that  this ansatz is a partial case of more general construction of
differential complex on
 $GL_q (2)$ [5,7].  Addition of the term $\xi(x)g(\tau)$ results in
a relation for the connection $A$ more complicated than (\ref{A17}) .

Consider the $q$-traceless curvature form for the connection (\ref{A29})
\be
F = \omega^2 f(\tau)[1-f(q^2 \tau)] + (q^2 \lambda)^{-1} \omega\xi
[f(\tau)-f(q^2 \tau)] \label{A30}
\ee

The anti-self-duality equation $\ast F=-F$ for our ansatz is equivalent
 to the nonlinear finite-difference equation
\be
f(\tau)-f(q^2 \tau)=(1- q^2 )f(\tau)[1-f(q^2 \tau)] \label{A31}
\ee

This equation has a simple solution  analogous to the classical
BPST-solution
\be
f(\tau)=\frac{\tau}{c+\tau}\;,  \label{A32}
\ee
where $c$ is an arbitrary constant. Note that our solution for
 connection  $A$  contains parameter $q$ only through definitions of
 $\omega(x)\;\mbox{and}\; \tau\; $  ,
however, the corresponding curvature has a more explicit $q$-dependence.

The curvature form can be written in terms of the field strength
\[
F= dx^i_\alpha dx^k_\beta F^{\beta \alpha}_{ki}(x)=d^2 x_{\alpha\beta}
F^{\beta \alpha} + d^2 x^{ik}F_{ki}\;,
\]
where Eq(\ref{A26}) is used.

The QGSD-equation for the field strength has the following form
\be
F^{\beta \alpha}_{ki}=\varepsilon_{ki}(q)F^{\beta \alpha} \label{A33}
\ee

It is interesting to discuss the $q$-deformation of the  harmonic (or
twistor)
formalism for QGSDE. The $q$-deformed harmonics can be considered as
 elements of $SU_q (2)$ matrix $u^i_a ,\;(i=1,\;2\;,\;a=+,\;-\;).$ We
 shall
 treat these matrix  elements as coordinates of the noncommutative
 coset space $SU_q (2)/U(1)$ by analogy with the classical harmonic
 formalism for a self-duality equation\cite{a16}.

Consider $SU_q (2)\times U(1)$ cotransformations of $q$-harmonics
\be
u^i_{\pm} \rightarrow l^i_k u^k_{\pm} \mbox{exp}(\pm i\alpha)\;,
\label{A34}
\ee
where $\alpha $ is $U(1) $parameter and $l$ is a matrix of left
$SU_q (2)$ acting on $E_q(4)$.

The $q$-harmonics satisfy the following relations:
\bea
R u_1 u_2 =u_1 u_2 R\;,& \;\;\;\;q u_1 x_2= R x_1 u_2 \;,&\\
\label{A35}
& \varepsilon_{ki}(q) u^i_{-}u^k_{+}=\sqrt{q} \;,& \nn
\eea

It is convenient to use the 3-dimensional left-invariant differential
calculus on $SU_q (2)\;\;\;$
 [9, 17] for the harmonic formalism. Consider the
$q$-traceless left-invariant 1-forms $\;\theta =S(u)du\;$ and introduce
 the notations:
\be
\theta_0=\theta_+^+=-q^{-2} \theta^-_-,\;\;\;\;\;\;\theta_{(+2)}=
\theta^-_+,\;\;\;\;\;
\theta_{(-2)}=\theta^+_-  \label{A36}
\ee

We shall below write  the left-invariant relations between
$\theta\;\mbox{and}\;u$ which allow us to define the operator of
 harmonic external derivative on $SU_q (2)$
\be
d_u = \delta_0 +\delta +\bar{\delta}=\theta_0 D_0 + \theta_{(-2)}
D_{(+2)} + \theta_{(+2)}D_{(-2)}\;, \label{A37}
\ee
where $\delta_0\;,\; \delta\; \mbox{and}\;\bar{\delta} $ are invariant
operators satisfying the ordinary Leibniz rules and the following
relations
\bea
& \delta_0^2 =\delta^2 =\bar{\delta}^2 =0 &\\ \nn
& \left\{\delta_0,\delta\right\}+\left\{\delta_0,\bar{\delta}\right\}
+\left\{\delta,\bar{\delta}\right\}=0 & \label{A38}
\eea

The left-invariant differential operators $D_0 ,\; D_{(\pm 2)}$ are the
basis
of a $q$-deformed Lie algebra  equivalent to the universal
enveloping
algebra ${\bf U}_q[SU(2)]$ \cite{a17}.

Let us define an invariant decomposition of the Maurer-Cartan equations
 for\\
 $SU_q (2)/U(1)$
\bea
& d_u \theta_0 =2\delta\theta_0=2\bar{\delta}\theta_0=-\theta_{(-2)}
\theta_{(+2)} &\\ \nn
& d_u \theta_{(+2)}=\delta_0\theta_{(+2)}=q^2 (1+q^2)\theta_0
\theta_{(+2)}
& \\ \label{A39}
& d_u \theta_{(-2)}=\delta_0\theta_{(-2)}=
 q^2 (1+q^2)\theta_{(-2)}\theta_0
& \nn
\eea

Define also $\delta_0\;,\; \delta \;\mbox{and}\;\bar{\delta}$ operators
on the quantum harmonics
\bea
 \delta_0 u^i_+=u^i_+ \theta_0=q^2 \theta_0 u^i_+\;,&\;\;\;
\delta u^i_+=0 &\\ \nn
& \bar{\delta} u^i_+=u^i_- \theta_{(+2)}=q^{-1}\theta_{(+2)}u^i_- &\\
\label{A40}
 \delta_0 u^i_-=-q^2 u^i_- \theta_0=- \theta_0 u^i_-\;,&\;\;\;
\bar{\delta} u^i_-=0 &\\ \nn
& \delta u^i_-=u^i_+ \theta_{(-2)}=q\theta_{(-2)}u^i_+ &
\label{A41}
\eea

Global functions on a quantum 2-sphere can be defined via the
invariant condition
$$
\delta_0 f(u)=\theta_0 D_0 f(u)=0
$$

Consider the harmonic decomposition of the Euclidean coordinates and
derivatives
\bea
 x_{(b)\alpha}=-q\varepsilon_{ik}(q)u^k_b x_{\alpha}^i\;,&\;\;\;\;
\partial^\alpha_a=u^i_a \partial^\alpha_i &\\ \label{A42}
& \partial^\alpha_a x_{(b)\beta}=\delta^\alpha_\beta \varepsilon_
{ba}(q)& \nn
\eea
where $a,\;b=+,\;-\;$.

One can use the asymmetric decomposition of the operator $d_x$ on
$E_q (4)$
\bea
& d_x =dx_{\alpha}^i \partial^\alpha_i = \bar{d} +(d_x-\bar{d}) &\\
\label{A43}
 \bar{d}\;\sim\; dx_{(-)\alpha}\partial^\alpha_+\;,&\;\;\;\bar{d}^2 =0
\;,&\;\left\{\bar{d},\delta\right\}=0  \nn
\eea
It should be remarked that the use of the symmetric decomposition
results in
a modification of analitycity condition for corresponding invariant
 operators.

An analyticity condition for the deformed harmonic space has the
following form
$$
 \partial^\alpha_+ \Lambda(x_{(+)},u)=0\;\;\Longleftrightarrow\;\;
 \bar{d} \Lambda=0
$$

Multiplying QGSDE (\ref{A33}) by $u^i_+ u^k_+ $ one can obtain the
$q$-deformed
integrability conditions in central basis (CB)( CB corresponds to
 $u$-independent
gauge-group matrices $T(x)\;$) which are analogous to the classical
self-dual integrability conditions [16 , 18].

Consider the decomposition of the $U_q (2)$-connection in CB
corresponding to
(42) and let $\bar{a}\sim dx_{(-)\alpha}A^\alpha_+ (x) $ be a
 connection
 for $\bar{d}$. The quantum-group self-duality equation\\
 (\ref{A33}) is
equivalent to the following zero-curvature equation
\be
\bar{d}\bar{a}-\bar{a}^2 = 0 \label{A44}
\ee

This equation has the following harmonic solution
\be
\bar{a}=\bar{d}h S(h)=\omega(h,\bar{d}h) \label{A45}
\ee
where $h(x,u)$ is a "bridge" $U_q (2)$-matrix function. The matrix
elements
of $h \;\mbox{and}\;\bar{d}h$ satisfy the relations analogous to
Eqs(12-15).
 Additional harmonic relations are
\be
\delta \bar{a}=0\;,\;\;\;\delta_0 h=0 \label{A46}
\ee

The bridge solution possesses a nontrivial gauge freedom
$$
h\;\to T(x)h\Lambda(x_{(+)},u)\;,\;\;\;\;\delta_0 \Lambda=0
$$
where $\Lambda$ is an analytical $U_q (2)$ gauge matrix.

The matrix $h$ can be treated as a transition matrix from the central
basis
to the analytical basis (AB) where $\bar{d}$ has no connection. The
characteristic feature of AB is  a nontrivial
 harmonic
connection $V$ that is an invariant component of a new AB-basis $A_{AB}$
 in the
algebra of $U_q (2)$ differential complexes
\bea
& A_{AB}= S(h)Ah - S(h)dh = \tilde{A} -S(h)d_u h = \tilde{A} + V &
 \\ \nn
& V=-S(h)\delta h -S(h)\bar{\delta} h= v + \bar{v} & \label{A47}
\eea
where the analytical connection $v=\theta_{(-2)} V_{(+2)}$ contains
the analytical prepotential $V_{(+2)}$.

By analogy with the classical harmonic formalism \cite{a16} the
 prepotential
 $V_{(+2)}$ generates a general solution of QGSDE that can be
 obtained
as a solution of the basic harmonic gauge equation
\be
\delta h + h v =0 \label{A48}
\ee

One can obtain   explicit or perturbative solutions of  this
equation
by using the noncommutative generalizations of classical harmonic
expansions
and harmonic Green functions [16 , 19]. It seems very interesting to
study reductions of QGSDE to lower dimensions and to search a more
general
deformation scheme for the self-duality equation.

{\bf Acknowledgments}

The author would like to thank V.P.Akulov, B.M.Barbashov, Ch. Devchand,
E.A. Ivanov, J.Lukierski,
V.I.Ogievetsky, Z.Popowicz, P.N.Pyatov, A.A. Vladimirov and especially
A.P.Isaev for helpful discussions and interest in this work.

I am grateful to administration of JINR and Laboratory of Theoretical
Physics for hospitality. This work was supported in part by
International
Science Foundation (grant RUA000) and Uzbek Foundation of
 Fundamental
Researches under the contract No.40.

\end{document}